\documentclass[fleqn,usenatbib]{rasti}

\usepackage{newtxtext,newtxmath}

\usepackage[T1]{fontenc}

\DeclareRobustCommand{\VAN}[3]{#2}
\let\VANthebibliography\thebibliography
\def\thebibliography{\DeclareRobustCommand{\VAN}[3]{##3}\VANthebibliography}

\usepackage{graphicx}	
\usepackage{amsmath}	
\usepackage{subcaption}
\usepackage{hyperref}

\title[MiraBest Dataset]{MiraBest: A Dataset of Morphologically Classified Radio Galaxies for Machine Learning}

\author[F.~A.~M. Porter \& A.~M.~M. Scaife]{
Fiona A.~M.~Porter,$^{1}$\thanks{E-mail: fiona.porter-2@manchester.ac.uk (FP)}
and Anna M.~M.~Scaife$^{1,2}$
\\
$^{1}$Jodrell Bank Centre for Astrophysics, Department of Physics \& Astronomy, University of Manchester, Oxford Road, Manchester M13 9PL UK\\
$^{2}$The Alan Turing Institute, Euston Road, London NW1 2DB, UK\\
}

\date{Accepted XXX. Received YYY; in original form ZZZ}

\pubyear{2015}

\begin{document}
\label{firstpage}
\pagerange{\pageref{firstpage}--\pageref{lastpage}}
\maketitle

\begin{abstract}
The volume of data from current and future observatories has motivated the increased development and application of automated machine learning methodologies for astronomy. However, less attention has been given to the production of standardised datasets for assessing the performance of different machine learning algorithms within astronomy and astrophysics. Here we describe in detail the MiraBest dataset, a publicly available batched dataset of 1256 radio-loud AGN from NVSS and FIRST, filtered to $0.03 < z < 0.1$, manually labelled by Miraghaei and Best (2017) according to the Fanaroff-Riley morphological classification, created for machine learning applications and compatible for use with standard deep learning libraries. We outline the principles underlying the construction of the dataset, the sample selection and pre-processing methodology, dataset structure and composition, as well as a comparison of MiraBest to other datasets used in the literature. Existing applications that utilise the MiraBest dataset are reviewed, and an extended dataset of 2100 sources is created by cross-matching MiraBest with other catalogues of radio-loud AGN that have been used more widely in the literature for machine learning applications. 
\end{abstract}




\section{Introduction}

In radio astronomy, morphological classification using convolutional neural networks (CNNs) and deep learning is becoming increasingly common for object classification, in particular with respect to the classification of radio galaxies \citep[see e.g.][etc.]{aniyan2017classifying,alger2018,wu2018,lukic2018,lukic2019,tang2019transfer,wang2021,ntwaetsile2021,bowles2021attention,sadeghi2021,e2cnn,becker2021,mohan2022,inigo2022}. Many of these works have focused on the morphological classification of radio galaxies following the Fanaroff-Riley classification scheme \citep[FR;][]{fanaroff1974morphology}, used to group radio-loud active galactic nuclei (AGN) by examining the locations of their regions of greatest luminosity relative to overall source extent. The initial scheme posited that there were two major populations of such sources - those which were core-brightened, with their peak luminosity concentrated at a radius of less than half than the overall angular size of the source from its centre (FR Type~I), and those which were edge-brightened, with their peak luminosity concentrated at a radius of more than half the angular size of the source (FR Type~II), and that there was a division in luminosity between the two populations at approximately $10^{25}$\,Watts\,Hz$^{-1}$\,sr$^{-1}$, with edge-brightened sources having a higher intrinsic luminosity than core-brightened sources. As described, this taxonomy requires that an AGN is associated with well-resolved extended emission external to the AGN core in order to be classifed as either FRI or FRII.

While the Fanaroff-Riley scheme was initially viewed as having a very straightforward luminosity boundary between morphological classes \citep{fanaroff1974morphology}, further study has shown that this is not the case, see (e.g.) \cite{hardcastle2020radio} for a review. In recent studies, sources have been detected which have raised questions about the use of this boundary; for example, \cite{mingo2019revisiting} found that around 20\% of FRII galaxies in their sample had radio luminosity below the traditional cutoff, some by as much as two orders of magnitude, meaning that there is some range of luminosities in which both classes can be found. As well as this, increasing survey sensitivity has allowed for the discovery of several classes with features that do not match the classic morphologies, including hybrid \citep[e.g.][]{gopal2000extragalactic, kapinska2017radio} and restarting sources \citep[e.g.][]{lara1999restarting, mahatma2019lotss}. Clearly, the dichotomy between Fanaroff-Riley classes is not only more complex than originally believed, but sufficiently complex that it is still not fully understood. While it is generally accepted that the same underlying mechanism likely powers all FR galaxies, with the different morphologies arising as a result of jet interactions with surrounding environments of different densities \citep{gopal2000extragalactic, kaiserbest, tchekhovskoy2016three, mingo2019revisiting, hardcastle2020radio}, the precise requirements of host galaxy characteristics, jet power, the properties of the inter-cluster medium and disruptions in that medium are still in question \citep{mingo2019revisiting, hardcastle2020radio}. To gain a stronger understanding of radio galaxies, it is necessary that we obtain more examples of FR sources. 

The new generation of radio surveys with telescopes such as LOFAR \citep{shimwell2017lofar, lotss}, MeerKAT \citep{jonas2016meerkat, Jarvis2016} and ASKAP \citep{Johnston2008, mcconnell2020rapid} are already providing additional data with which to expand our catalogues of radio galaxies, as does the advent of surveys with the Square Kilometre Array (SKA) telescope \citep{braun2015advancing, grainge2017square}. While the unprecedented depth of field and resolution will permit the detection of sources that previously could not be observed, the volume of data produced by the SKA and its precursors will be such that robust machine learning classification will be absolutely vital to allow sources of interest to be identified and classified. 

Currently, the archival data sets available for training radio galaxy classifiers are of comparable size to many of those used in computer vision \citep[e.g. CIFAR;][]{krizhevsky2009}, with around $10^5$ samples available. However, a fundamental difference is the degree of domain knowledge needed for creating \emph{labelled} data sets, which has a much higher cost for radio astronomy data. As a result of this, labels are sparse in radio galaxy data sets, with labels only included for a small fraction of data. Indeed, the majority of existing catalogues (and hence labelled datasets) of FR galaxies contain only a small number of sources - typically several hundred at most \citep{gendre2008combined, capetti2017fricat, capetti2017friicat, koziel2020radio}, with \cite{mingo2019revisiting} as a recent exception. Consequently, when constructing a machine learning dataset, the use of the largest practical number of images of each class is preferable, ensuring that the full variability of features within each class is captured by the dataset; without this, there is a risk that any models using that dataset will have a limited ability to generalise to images which show features that are not adequately represented within it. 

In this work we describe the MiraBest batched dataset\footnote{The MiraBest dataset can be downloaded from: \url{https://doi.org/10.5281/zenodo.4288837}.}, a publicly available FR-labelled machine learning data set of radio galaxies. With 1256 samples, MiraBest is currently the largest publicly available machine learning dataset for radio galaxy classification. The structure of this paper is as follows: in Section~\ref{sec:datasets} we outline the general principles underlying the construction of astronomical machine learning datasets; in Section~\ref{sec:selection} we describe the sample selection and dataset structure for MiraBest; in Section~\ref{section: processing methods} we describe the pre-processing applied to MiraBest data samples, and in Section~\ref{section: composition} we provide an analysis of the overall dataset composition; in Section~\ref{section: comparisons} we compare the MiraBest dataset to other radio galaxy machine learning datasets in the literature, and in Section~\ref{sec:supplementary} we describe additional supplementary datasets that are provided with the core MiraBest dataset; in Section~\ref{sec:literature} we outline existing applications of the MiraBest dataset in the literature and in Section~\ref{sec:conclusions} we draw our conclusions.

\section{Constructing datasets of astronomical sources}
\label{sec:datasets}

While a large number of astronomical catalogues are accessible for general use, not all are suitable to be used to create machine learning datasets. Astronomical catalogues are constructed to serve a variety of specific purposes; depending on the field that seeks to collate them, which properties are considered useful and which are irrelevant can vary significantly, and a scientifically useful catalogue may lack the features needed to produce a useful machine learning dataset. In the case of a dataset intended for source classification with supervised learning, the following properties should be considered when attempting to build a dataset from an existing catalogue.

\subsection{Number of sources}

An obvious area in which machine learning datasets and and astronomical catalogues may differ is size. A catalogue of a few dozen rare astronomical objects may be enough to glean a number of astrophysical properties and constraints, and allow astronomers to identify methods by which they might be able to find more of these rare objects, see e.g. \cite{lyne2017two}, \cite{hartley2017}, \cite{titus2020radio}, \cite{pleunis2021fast}, \cite{rezaei2022}; indeed, the small number of sources available might allow for each to be studied in more depth, see e.g. \cite{young2013long}, \cite{fermi2015extremely}, \cite{chime2020periodic}. A machine learning dataset of a dozen images, however, is of very dubious use; individual classes within a dataset being this small has been shown to result in poor classification ability \citep{cho2015much}, and while there is no definitive answer for the minimum quantity of data required for a machine learning model, a suggested rule of thumb is that the number of samples in a dataset should be at least a factor of fifty larger than the number of classes \citep{alwosheel2018your}. Machine learning datasets hence have a much higher minimum population requirement to produce ``good" science - a dataset of several hundred images is considered very small, and while data augmentation can artificially increase the size of a small dataset, making it more likely to be useful, a larger quantity of unique data is significantly more likely to produce accurate and generalisable results than a smaller augmented dataset \citep{brigato2021close}.

\subsection{Availability of labelled data}

A vital component of a machine learning dataset intended for supervised learning is that all sources included within it must be labelled accurately, and a lack of appropriately labelled data is accordingly a significant issue \citep{raghu2020survey}. Manual classification to create an appropriately-sized dataset is time-consuming and requires sufficient knowledge of the classes involved that it is ensured they are labelled correctly; depending on the type of source involved, even multiple human classifiers with expert knowledge might not agree on what the correct class is for a particular object \citep{nair2010catalog, walmsley2022galaxy}, especially if the source population is small enough that there are relatively few examples of any given class available for comparison. For this reason, a number of astronomical machine learning datasets draw from previously-created source catalogues rather than their creators seeking out and labelling entirely new sources \citep[e.g.][]{aniyan2017classifying, tang2019frdeep}, including MiraBest.

In addition to this, machine learning datasets require some wariness towards ``label noise" - that is, the uncertainty introduced into a machine learning model as a result of some of the images in a dataset being incorrectly labelled \citep[see e.g.][]{frenay2014classification}. On the whole, this is not typically a problem in other astronomical catalogues, as a small number of atypical sources are not usually expected to have a significant impact upon results derived from a large population; however, it can negatively impact the training of a machine learning model, as the inclusion of mislabelled images can result in the model being effectively ``taught wrong" with regard to which features are associated with which class \citep{northcutt2021pervasive}. \cite{miraghaei2017nuclear} has the unusual distinction of not only providing class labels but information on whether the authors were confident in their classification, effectively providing a measure of uncertainty on the labels themselves and allowing for the potential to study how models respond to sources for which astronomers are unsure of the appropriate morphological class.

\subsection{Inclusion of  ``messy" data}

In general, when collating a dataset for astronomical study, there is an expectation that the data contained within will be ``good data" - that is, as free as possible from artefacts, with any effects from background sources removed, etc. - and that the properties observed are exclusively from the intended source. In machine learning, though, there is merit in ``bad" or ``messy" data being included: it reflects the astrophysical reality of the environments of many sources \citep{norris2021evolutionary}. While artefacts might render an observation not terribly useful for determining precise properties of a source, or background sources might make it difficult to determine which emission can be attributed to which object, it is plainly apparent that these will occur in survey data nonetheless; it is hence better, when using machine learning, to have developed a model that can identify the features necessary to classify a source despite artefacts, or which has learned that background emission may be present without affecting the class of the primary source, than to use one which has never encountered these circumstances and cannot adapt to them without retraining. Focusing on ``good" data not only reduces the possible size of a machine learning dataset: it limits its ability to recognise anything but other ``good" data, which is a particular issue when ``messy" data can be more often representative of unusual features that would warrant additional attention from astronomers \citep{norris2021evolutionary, gupta2022discovery}.

This also applies to the inclusion of atypical sources for a given class, such as rare subclasses; while these may be considered to add a form of ``noise" to a dataset when their properties noticeably differ from the ``standard" population, exclusively using sources with ``standard" morphology in machine learning datasets again may limit the ability of the resulting models to recognise unusual sources within survey data. While there may not always be a sufficient number of examples of a given rare subclass to allow a model to specifically recognise that subclass’s properties, there is some merit in including such sources in a dataset in a manner that lets them easily be screened out or merged into a larger class; it is far easier to remove unwanted images from a machine learning dataset than it is to add new images that perfectly match the original's data processing methods, and these sources may be used to examine model responses to unusual objects which are nonetheless part of the same overall source population.

\section{Selection method and dataset structure}
\label{sec:selection}

The sources classified by \cite{miraghaei2017nuclear} were identified from a catalogue of radio-loud AGN \citep{best2012fundamental}, which had cross-matched galaxies from the seventh data release of the SDSS \cite{sdssdr7} with radio components from NVSS \citep{NVSS} and FIRST \citep{first} surveys conducted at 1.4\,GHz with the Very Large Array (VLA) telescope. This catalogue was filtered to obtain sources that had a lower redshift limit of $z > 0.03$, as the large angular size of nearer galaxies was deemed likely to result in catalogued SDSS parameters containing errors, and an upper limit of $z < 0.1$ to ensure spectroscopic classification would be available. A flux density cut of 40\,mJy was applied to ensure that there was a signal to noise ratio large enough to detect diffuse radio emission.

To obtain a Fanaroff-Riley class, images of each source from either NVSS or FIRST were inspected visually and labelled using the original definition of the two classes \citep{fanaroff1974morphology}: whether the distance from the AGN's centre to the regions of brightest emission on either side was less (FRI) or more (FRII) than half of the angular extent of the radio emission. 

Due to the limitations of FIRST and NVSS's angular resolution and ability to detect faint emission, some objects could not be labelled with confidence; as a result, these sources were flagged as "uncertain" in label. If a galaxy was noted to show a non-standard FR morphology, this was likewise flagged. In addition to the standard FRI and FRII sources, a small number of sources were classified as hybrids - showing morphology characteristic of FRIs on one side and of FRIIs on the other - or ``unclassifiable", showing no obvious FR morphology.

Each source within \cite{miraghaei2017nuclear}'s catalogue was characterised by a three-digit class label. These denote, in order, overall Fanaroff-Riley class, degree of confidence in classification, and morphological subclass. Not all possible combinations of digits are present within the catalogue; for example, double-double morphology \citep{schoenmakers2000} is characteristic of FRII sources, so no FRI sources exist with this label. As hybrid sources have no real ``standard" or ``non-standard" morphologies, all hybrid sources are deemed to be morphologically ``standard". The meaning of each digit's possible values are listed in Table~\ref{tab: class labels}.

\begin{table}
    \centering
    \begin{tabular}{lll}
        \hline
         Digit 1 & Digit 2 & Digit 3  \\\hline
         1 - FRI    & 0 - Confident   & 0 - Standard \\
         2 - FRII   & 1 - Uncertain   & 1 - Double-double \\
         3 - Hybrid &                 & 2 - Wide-angle Tail \\
         4 - Unclassifiable &         & 3 - Diffuse \\
                    &                 & 4 - Head-tail \\\hline
         
    \end{tabular}
    \caption{The three-digit classification scheme used by \citet[][]{miraghaei2017nuclear}. Double-double sources are exclusively FRII; wide-angle, diffuse and head-tail sources are exclusively FRI.}
    \label{tab: class labels}
\end{table}

While the catalogue lists 1329 sources, not all were used within the MiraBest dataset. A total of 73 sources were excluded for the following reasons:

\begin{enumerate}
  \item 40 sources were labelled as ``unclassifiable" within \cite{miraghaei2017nuclear}; while it is not stated why they were deemed unclassifiable, they are assumed to show no recognisable FR morphology and hence provide no information for the classification of FR galaxies. While they do represent a population of non-FR sources that might be encountered while surveying, this population is too small to be useful within the dataset as an inbuilt ``non-FR'' class; as such, they were removed.
  \item 28 sources were found to have an angular extent greater than 270 arcseconds. As images within the dataset were processed to have dimensions of 150 x 150 pixels, corresponding to 270 arcseconds in the FIRST survey, any sources larger than this were removed to prevent the use of partial sources being present in the dataset.
  \item 4 sources were located partially or fully outwith the area of sky covered by the FIRST survey. As significant portions of each source lacked image data, they were removed to prevent the use of partial sources.
  \item 1 source, with image label 103 (confidently-labelled FRI with diffuse morphology), formed a single-source subclass. As a minimum of two images per subclass were required for inclusion in the dataset - one for the training set and one for the test set - this image was removed.
\end{enumerate}


With these sources removed, the remaining 1256 sources were used to create the MiraBest dataset.

\section{Processing methods}
\label{section: processing methods}

For the purposes of allowing direct comparison to and possible combination with an already extant dataset of Fanaroff-Riley galaxies - FR-DEEP, presented by \cite{tang2019transfer} - the data processing used was matched closely to this dataset's methods. Further information on this dataset and its processing techniques may be found in Section~\ref{section: frdeep}.

Data from the FIRST survey were obtained as .fits files using the SkyView Virtual Observatory \citep{skyview} for a $300\times300$ pixel area of sky centred upon each set of coordinates provided by \cite{miraghaei2017nuclear}, equivalent to an angular extent of 540". A standardised naming system for each source was implemented at this stage, with a multiple source properties stored within the name string using the following format: ``[3-digit class label]\_[source right ascension]\_[source declination]\_[redshift]\_[angular size]", with right ascension and declination given in decimal degrees and angular size in arcseconds. These properties were stored within the file names to ensure that source coordinates and labels could be rapidly and easily matched to their respective images. 

\subsection{Noise reduction}

Image data from FIRST \citep{first} as provided by SkyView \citep{skyview} at this stage contained sufficient radio noise that the central FR source was not always readily apparent from visual inspection. To reduce this radio noise, sigma-clipping was performed using the \verb|astropy| package's \verb|sigma_clipped_stats| function \citep{astropy2022astropy}. Any pixel that was found to have a radio flux density less than $3\sigma$ from the image's mean was set to zero; this process was repeated a maximum of five times, stopping early if no pixels below the clipping threshold were found. Performing sigma-clipping at this stage - with an image size greater than that which would be used for the final dataset - allowed for the most complete possible removal of noise without loss of source information, as providing a larger quantity of noisy pixels allowed for better characterisation of the radio background. At this stage, images were considered to have been ``cleaned", and were cropped to $150\times150$ pixels centred upon the radio galaxy.

\subsection{Removal of extraneous data and background sources}
\label{section: removing background}

The maximum angular size of sources used in this dataset was 270", corresponding to 150 pixels in FIRST. However, as the images created were square, not circular, ``unwanted" data was present at the corner of each image that was not expected to have any relevance to the FR source. To ensure that these regions would not provide any possibly confounding data, a circular mask was applied to each image. Initially, it was considered whether it might be useful to customise the size of this mask to each image, masking out any data beyond the angular size of the source stated in \cite{miraghaei2017nuclear}, to completely remove any possible bright background sources. However, this option was rejected for the following reasons:

\begin{enumerate}
  \item Using masks set to the angular sizes provided by \cite{miraghaei2017nuclear} frequently resulted in pixels that clearly contained source data to be masked out. This discrepancy between the stated and observed extents of the sources might be a result of the authors using NVSS data to determine angular size, as NVSS's 15" per pixel resolution could readily result in single-pixel underestimations of size that correspond to multiple pixels at FIRST's 1.8" resolution. Correctly setting the mask limits would hence require visual inspection of every image to be used in MiraBest to determine the appropriate size for its mask, a task which would be time-consuming to complete because the asymmetric structure and diffuse emission of many sources were found to require testing multiple sets of limits per image to ensure that no emission was mistakenly masked out. Given that relatively few images clearly benefited from this more curated approach to masking, it was deemed to be an inefficient use of time to do so.
  \item While bright background sources might somewhat impact the performance of machine learning classifiers on this dataset, the presence of such sources is astrophysically normal; it is not only possible but inevitable that some images produced by future radio telescopes will contain background sources within the field of view of a target source, and it is unrealistic to expect that they will all be cleaned from these images before classification is attempted. For this reason, radio background sources within the central 270" of an image represent behaviour that will likely be seen in new astrophysical images, and their inclusion was deemed to be unlikely to harm the usefulness of this dataset as a whole.
\end{enumerate}
    
Because of this, rather than a variable mask, a fixed circular mask of diameter 150 pixels was applied to every image to remove data that was unambiguously not associated with the FR source.

\subsection{Normalisation}
\label{section: normalisation}

The images included in MiraBest at this stage naturally showed variation in source flux density; FR galaxies are not standard candles, so even sources at the same redshift can show markedly different maximum flux densities, and the population used in MiraBest covers redshift range $0.03 < z < 0.1$, resulting in more distant galaxies tending to be fainter. While, as with the presence of background sources, variation in source flux density is inevitable in future surveys, a potential issue caused by inherent properties of FR galaxies was noted if the images were not normalised.

FRII galaxies are known to be on average intrinsically brighter than FRIs \citep{mingo2019revisiting}. A machine learning model trained on images which preserve the innate flux densities might result in the model, rather than identifying morphology, basing much of its classification on maximum flux density alone. Such a model could be expected to be easily confounded when applied to unseen images from surveys with greater sensitivity than FIRST; it might be expected to simply label all images below a certain flux density as being FRIs, rendering it useless for surveys besides FIRST.

To prevent this from occurring, all images were normalised. This was done by identifying the minimum and maximum flux density values in each image, and scaling each pixel's value as follows:

$$Normalised\,pixel\,value = 255 \times \frac{Pixel\, value - minimum\, flux}{Maximum\, flux - minimum\, flux}$$

The factor of 255 is used to facilitate image conversion into PNG files; greyscale PNGs have pixel values ranging from 0 (darkest) to 255 (brightest), so multiplying by this factor ensures maximum dynamic range by setting the minimum pixel value to 0 and the maximum to 255. At this stage, all image processing is complete; all images are converted to PNG format and are ready to be collected into a dataset.

PNG files are here favoured over retaining the original FITS format to make this dataset accessible to non-astronomers who wish use astronomical data for machine learning. Although PNGs have a more limited dynamic range than those of FITS files, the FR classification scheme is primarily concerned with the location of the brightest regions of a source, making the potential loss of very faint emission from these sourcse unlikely to affect the ability to classify images in this case. As the source data was originally in monochrome 1.4\,GHz flux, conversion into single-channel greyscale PNG is able to represent relative flux within an image, and has the additional benefit of reducing typical file size by two orders of magnitude, making the dataset more practical to download and store if desired.

\subsection{Constructing the batched dataset}

MiraBest's images are $150\times150$ pixels, making them relatively large when compared with commonly used benchmarking datasets such as MNIST  \citep[$28\times28$ pixels;][]{lecunmnist} or CIFAR10  \citep[$32\times32$ pixels;][]{cifar10}, although we note that Imagenet \citep{imagenet} contains a variety of image sizes, including those that exceed MiraBest. However, even when using three channels to provide RGB information rather than MiraBest's single greyscale channel, a MiraBest image still contains around an order of magnitude more data than CIFAR10. For this reason, it was designed for use as a \emph{batched dataset} \citep{masters2018revisiting}; loading each batch into memory sequentially is less computationally demanding than loading the entire dataset at once, making MiraBest practical for use with devices with relatively small memory, such as personal laptops. It was decided to separate the dataset into eight batches of 157 images: seven were to be labelled as training set batches, with the last reserved to be a test set batch.

The classes within MiraBest are not balanced; that is, there is not an equal number of sources in every class (see Section~\ref{section: composition} for a full discussion of the class breakdown). While in a balanced dataset, it is possible to randomly separate all images into different batches and expect a reasonably equal quantity of each class to be present in each batch, this is not the case for a dataset that, like MiraBest, has a number of subclasses that represent a very small proportion of the whole. Randomly selecting images in this case might result in some batches completely lacking in particular sources, which is a particularly pressing concern for a dataset's test batch; if the test set contains no examples of a particular class, there is no way to evaluate a machine learning's model's performance on that class during training, and the model risks significant overfitting as a result.

To prevent this from occurring, a fixed batch structure was used to ensure that a roughly equal quantity of sources of each class were present in every batch. The source quantities present in \cite{miraghaei2017nuclear} were such that it was impossible to ensure exactly identical composition between all batches, but the following method was used to ensure that sources were distributed as evenly as possible:

\begin{enumerate}
  \item The total number of images in each class was divided by eight and rounded down to determine the base number of images per class to include in each batch.
  \item Images were shuffled, and the base number of images of each class were assigned to each batch. If fewer than eight images of a class were available, one was randomly chosen to be reserved for the test batch to ensure that there were no instances of a class missing a test set image.
  \item The number of images in each batch and quantity of remaining unassigned images in each class were determined.
  \item Beginning with the test batch, each batch was iteratively filled with unassigned images. Each iteration, the class with the largest number of remaining images was determined, and an image of that class was assigned to the batch until it reached a total of 157 images; if all classes had an equal number of unassigned images, a class was selected at random. This method ensured that the most populous classes were preferentially added to the test batch, with the rarer subclasses more likely to be present in one of the training batches.
\end{enumerate}

Once the composition of the batches was determined, the file names of the images used in each batch were saved. This was done to allow for direct comparison between any datasets created using the same catalogue but using data from a different survey; ensuring the composition is entirely consistent prevents any potential differences in behaviour caused by having different proportions of sources in the different batches.

With batch structuring complete, the image data and labels for the sources within in each batch were collected to form the final dataset.

\section{Dataset composition and analysis}
\label{section: composition}

The overall composition of the MiraBest dataset is detailed in Table~\ref{tab: mirabest structure}.
\begin{table*}
\begin{tabular}{llllll}
\hline
Class & No. of images & Confidence & Morphology & No. of images & Class label \\ \hline
FRI & 591 & Confident & Standard & 339 & 0 \\ \cline{4-6} 
 &  &  & Wide-angle tail & 49 & 1 \\ \cline{4-6} 
 &  &  & Head-tail & 9 & 2 \\ \cline{3-6} 
 &  & Uncertain & Standard & 191 & 3 \\ \cline{4-6} 
 &  &  & Wide-angle tail & 3 & 4 \\ \hline
FRII & 631 & Confident & Standard & 432 & 5 \\ \cline{4-6} 
 &  &  & Double-double & 4 & 6 \\ \cline{3-6} 
 &  & Uncertain & Standard & 195 & 7 \\ \hline
Hybrid & 34 & Confident & Standard & 19 & 8 \\ \cline{3-6} 
 &  & Uncertain & Standard & 14 & 9 \\ \hline
\end{tabular}
\caption{The population of all FR classes within the MiraBest dataset, including the internal class labels used.}
\label{tab: mirabest structure}
\end{table*}
The three-digit class labels used by \cite{miraghaei2017nuclear} were reduced to single-digit class labels within the dataset to match conventions with other machine learning datasets. As these single-digit labels are less informative at a glance of an image's class, the three-digit method will be preferentially used going forward to ensure clarity. In addition to FR class, each source's right ascension and declination can be retrieved from the image's filename, making them fully traceable. While this is not typical for machine learning datasets in general, it is a useful measure for an astronomical dataset as this allows for easy cross-matching of sources between different catalogues.

\subsection{Dataset analysis}

With 1256 sources, MiraBest currently represents the largest publicly available machine learning dataset of Fanaroff-Riley galaxies. It is also the only known dataset that provides examples of clearly labelled non-standard morphology FRs, and hence contains not only the largest quantity but also the greatest morphological variety of FR galaxies. While at present some morphologies are represented by only a few samples, it can be expected that their numbers will significantly increase with wide-field, sensitive radio surveys, and their presence within the overall population of FR galaxies should not be neglected.

When considering broad FR morphology, there is a mild class imbalance between FRI and FRII sources, with forty more FRII sources than FRIs. However, this is considered unlikely to result in any noticeable effects upon performance of machine learning models; while significant class imbalances can lead to a model learning it can obtain a high overall accuracy by labelling all or most images as the majority class \citep{johnson2019survey}, the ratio of binary FRI/FRII sources here is approximately 48/52, which is not a large enough discrepancy to expect this behaviour.

When considering individual morphological subclasses, a much more significant imbalance can be observed. The most populous subclass, confidently-labelled wide-angle tail FRIs (class 102), is less than 15\% the size of its standard-morphology counterpart (class 100), and less than one in one hundred FRIIs included shows double-double morphology (class 201). As a result, while their inclusion as a whole benefits MiraBest by showing examples of less common morphologies that are nonetheless part of the Fanaroff-Riley classification system as a whole, this dataset is not well-suited to be used to train machine learning models intended to specifically identify unusual morphologies; there are simply not enough examples of each class for a model to be able to learn to classify them without severe overfitting, even with data augmentation. For this reason, MiraBest will generally better serve the needs of the astronomical community if the morphological subclasses are grouped with the overall population of their FR class; see Section~\ref{section: derivative datasets} for discussion of the merits of this approach and the derivative datasets that have been created for this purpose. 

The hybrid FR galaxies included in MiraBest likewise represent a very small portion of the dataset. Again, such a drastic class imbalance renders these images unlikely to be useful if attempting to develop a machine learning model that can differentiate FRI, FRII and hybrid FR sources, even with data augmentation; instead, they may best be used in identifying ways in which hybrid sources confound binary FRI/FRII classifiers \citep{mohan2022}. Even so, the quantity of images available may not allow for a large enough sample to be statistically significant, and for this reason a separate, larger dataset of hybrid sources was created, incorporating the sources in MiraBest; see Section~\ref{section: hybrid set} for information about this dataset and the catalogues of hybrid FR sources it draws from.

\begin{figure*}
        \centerline{
        \includegraphics[width=0.4\textwidth]{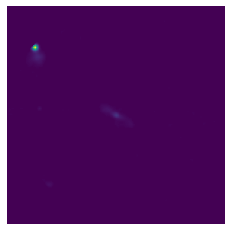}\qquad \includegraphics[width=0.4\textwidth]{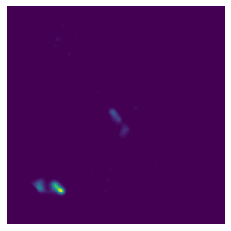}
        }
        \centerline{(a) \hskip 0.4\textwidth (b)}
        \vskip .1in
        \centerline{
        \includegraphics[width=0.4\textwidth]{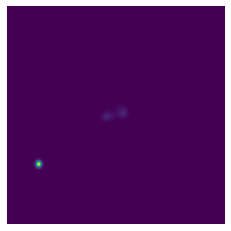}\qquad \includegraphics[width=0.4\textwidth]{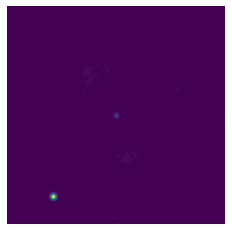}
        }
        \centerline{(c) \hskip 0.4\textwidth (d)}
        \caption{Examples of MiraBest images where the brightest source visible is not the central FR source. (a) Confidently-labelled FRI (100); (b) Uncertainly-labelled FRI (110); (c) Confidently-labelled FRII (200); (d) Uncertainly-labelled FRII (210). Although faint, the FR sources remain visible despite the presence of a background source affecting the normalisation.}
        \label{fig: background sources}
\end{figure*}

Following the decision not to remove background sources within a 135" radius of the central FR sources, there are several images in which the FR source is comparatively faint, with a bright background source having a noticeably brighter radio flux density; a selection of these images are shown in Figure~\ref{fig: background sources}. As discussed in Section~\ref{section: normalisation}, while this does introduce some noise into the dataset by including data that is unrelated to FR galaxies, background sources like these are expected to be present in images from other surveys, and any machine learning model that is to be used on data from a new survey will need to be able to identify FR galaxies whether or not a background source is present. Examples of bright background sources being included in MiraBest, then, might serve to make more robust classifiers by including these small pieces of irrelevant data.

\begin{figure*}
\centerline{
\includegraphics[width=0.3\textwidth]{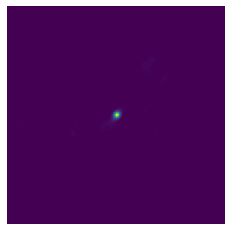} \qquad \includegraphics[width=0.3\textwidth]{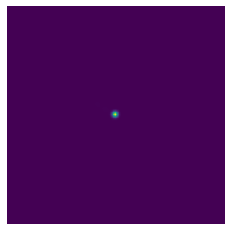} \qquad \includegraphics[width=0.3\textwidth]{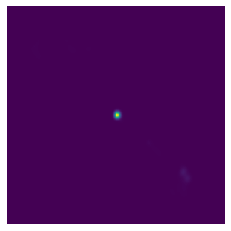}
}
\centerline{ (a) \hskip 0.3\textwidth (b) \hskip 0.3\textwidth (c) }
    \caption{Three examples of FR galaxies that are observed to have point-like morphology. (a) Confidently-labelled FRI (100); (b) Uncertainly-labelled FRI (110); (c) Confidently-labelled FRII (200).There is no clear sign of emission having been lost due to sigma-clipping, so it is assumed that these sources simply have a small enough angular size that obvious FR features are not apparent.}
    \label{fig: pointlike sources}
\end{figure*}

MiraBest uses any source from \cite{miraghaei2017nuclear} with an angular size less than 270"; as a result, some sources included in the dataset have very small angular sizes, and five are visually very close to point-like. Of these five, four are FRI sources, and three of them are uncertainly labelled. One point-like source of each class is shown in Figure~\ref{fig: pointlike sources}. FRIs are expected to be much more likely than FRIIs to present this morphology, as their core-brightened emission can result in a source with small angular size compared to the resolution of the telescope used having only their central region detected, with diffuse jets being undetected. It is not apparent that the sigma-clipping process has removed noticeable amounts of source emission from these images, so it is assumed that these images are accurate representations of the radio emission of these sources. While, to humans, morphological information might be difficult to glean from these images, it remains possible that a machine learning model might be able to identify some properties of these sources that allow it to classify them accurately, and perhaps be capable of distinguishing them from other point-like radio sources; for this reason, and because they represent a very small proportion of the dataset overall, they were retained. 

While MiraBest is considerably smaller than many machine learning datasets, this is unavoidable considering the comparative rarity of FR galaxies. As additional sources are identified that appear within the extent of the FIRST survey, it will become possible to extend it further to allow for a greater representation of the entire FR population; meanwhile, however, it remains the most comprehensive known machine learning dataset of FR galaxies. For direct comparison between MiraBest and other datasets of FR galaxies, see Section~\ref{section: comparisons}.

\subsection{Derivative datasets}
\label{section: derivative datasets}

As can be seen in Table~\ref{tab: mirabest structure}, the unusual morphological subclasses represent a small proportion of the entire dataset; ``standard" FRI and FRII sources represent approximately 92\% of the images, with most other morphologies comprising too small a proportion for their behaviour to be considered representative of their subclass as a whole. For this reason, additional class wrappers were created to allow the classes to be reduced down to simply FRI and FRII populations. The resulting simplified datasets were produced:

\begin{table*}
\begin{tabular}{lllllllc}
\hline
Name & No. of images & FR classes & Confidence & No. of images & Class label & Best reported accuracy & ref. \\ \hline
MBFRFull & 1222 & FRI & Any & 591 & 0 & 86.9 $\pm$ 0.5\% & [1]\\ \cline{3-6} 
 &  & FRII & Any & 631 & 1 & &\\ \hline
MBFRConfident & 833 & FRI & Confident & 397 & 0 & 96.54 $\pm$ 1.29\% & [2]\\ \cline{3-6} 
 &  & FRII & Confident & 436 & 1 & &\\ \hline
MBFRUncertain & 389 & FRI & Uncertain & 194 & 0 & N/A &\\ \cline{3-6} 
 &  & FRII & Uncertain & 195 & 1 & &\\ \hline
\end{tabular}
\caption{The three derivative datasets created from MiraBest with simplified internal class labelling system and the highest classification accuracy reported in the literature for each, where [1] \citet[][]{inigo2022}; [2] \citet[][]{e2cnn}.}
\label{tab: mirabest derivatives}
\end{table*}

Reducing the labelling system to these simplified variants allows the study of FR galaxies to be reduced to a binary classification problem, removing the possibility that an image could be considered to be ``misclassified" if a machine learning model predicts the correct FR class but a different human confidence label or morphological subclass than is given by the dataset; as humans, we would recognise the former as being irrelevant to FR class and the latter to be ``broadly correct'', but this distinction would not be made by a machine learning model by default. 

The possibility of some misclassifications being ``less wrong" than others in this way leads to confusion in the overall ability of a model to classify FR sources accurately, and with such small populations of sources with unusual morphology being available, it is often more useful to simply group all sources of a particular FR class together to better be able to analyse the population as a whole. While the internal labels are simplified, it is of note that neither the morphological information nor the confidence of any image are lost - these are independently accessible via the image filename, and thus it remains possible to identify any subset of interest within the dataset as desired.

\section{Supplementary hybrid dataset}
\label{sec:supplementary}

Hybrid radio galaxies are not truly out-of-distribution sources, given that the same underlying mechanism is believed to create both hybrids and binary FR sources, but they nonetheless represent a sources of confusion for machine learning models trained on binary sources; they could viably show properties of both classes simultaneously, resulting in models finding them equally probable of belonging to both binary classes. Hybrid galaxies also present an interesting population to study; their discovery helped to support the theory that FR morphology is at least part environmentally driven \citep{gopal2000extragalactic}, but the mechanism behind their formation is still not fully understood and it has been questioned whether they truly represent a separate FR class \citep[see e.g.][]{stroe2022host}. In order to identify these sources for study, it is important to be able to separate them from other radio galaxies and ensure they are not accidentally misidentified as binary FR sources.

\subsection{Constructing the hybrid dataset}
\label{section: hybrid set}

The number of hybrid FR sources contained in the catalogue used to create MiraBest is not particularly large; there are only 34 sources with hybrid labels, 19 being confidently labelled and 15 uncertainly labelled, while the binary FR classes have many hundreds of examples. To provide a more reliable analysis of the greater population of hybrid FRs, additional hybrid sources were sought out. As this variety of FR morphology is seen far less frequently than binary FRs, it was not expected that it would be possible to identify an equally large population, but even a slight increase in the number of hybrid sources would allow for a more statistically thorough exploration of their properties.

Two additional catalogues of hybrid sources were identified; \cite{kumari2021search} searched systematically within FIRST data to locate 45 confirmed hybrids and 5 candidate hybrids, while \cite{kapinska2017radio} located 25 candidate hybrids within FIRST via the Radio Galaxy Zoo citizen science project \citep{rgz}. These were assessed for suitability for inclusion in a hybrid dataset matching the image processing methods of MiraBest.

Sources in these two catalogues were labelled as either hybrid or hybrid candidate rather than MiraBest’s confident and uncertain classification labels; to allow for a unified labelling system, the former class was taken to be equivalent to the confidently classified hybrids in MiraBest and the latter equivalent to uncertainly classified hybrids. The list of coordinates was compared to those of the hybrids in MiraBest to check for duplicate sources, resulting in two sources from \cite{kumari2021search} being removed for being already contained in MiraBest.  One of these was labelled as a confident hybrid by Miraghaei \& Best, while Kumari \& Pal listed it as a hybrid candidate, highlighting that even when classified by humans, a degree of uncertainty remains in classification confidence; the second was agreed by both to be a certain hybrid. 

Two sources from \cite{kapinska2017radio} were found to have angular sizes exceeding the 270” size limit imposed upon MiraBest, and were hence removed from consideration; upon processing, a third was identified visually to have some of its emission cropped by the image processing, and it was likewise removed. Of the remaining images, seven were noted to have a bright source present that did not appear to be part of the hybrid emission. While it would be possible to remove any background sources that are clearly separated from the hybrid galaxy, the presence of background sources is not unexpected within radio data, and thus allowing these images to remain with background sources intact provides a more realistic representation of the environments around hybrid galaxies. 

With this, a dataset of 104 hybrid sources was created, with a total of 63 confident and 41 uncertainly labelled hybrids; 34 sources from the original set included in MiraBest, 48 from Kumari \& Pal, and 22 from Kapińska et al. This represents a threefold increase upon the number originally provided by MiraBest, and is believed to be the largest single dataset of hybrid FR sources available at present.

\section{Comparison to existing FR datasets}
\label{section: comparisons}

As discussed previously, the population of labelled Fanaroff-Riley galaxies is relatively small, and consequently relatively few datasets have been constructed focusing upon these sources. Most derive their sources from a combination of several FR catalogues, notably FRICAT \citep{capetti2017fricat} and FRIICAT \citep{capetti2017friicat} (both catalogues of FR galaxies found within the FIRST survey), and CoNFIG (the Combined NVSS-FIRST Galaxies sample) \citep{gendre2008combined}. Some background regarding the composition of these catalogues will be discussed before examining the datasets using these sources.

\subsection{FRICAT and FRIICAT}
\label{section: fricat}

Both FRICAT and FRIICAT were developed with the intention to provide large and comprehensive catalogues of Fanaroff-Riley galaxies that would allow for better understanding of their overall properties, with the particular goals of allowing their luminosity functions, environment, and evolution to be studied \citep{capetti2017fricat, capetti2017friicat}. At the time of its creation, FRICAT was the single largest catalogue of FRI galaxies in existence, containing more than double the sources of previous catalogues.

Both catalogues draw from a parent sample of radio sources visible in FIRST from \cite{best2012fundamental} that provide information regarding whether their emission is associated with an active galactic nucleus. They limit their sample for consideration to those with redshifts $z < 0.15$ - greater than \cite{miraghaei2017nuclear}'s limit of $z < 0.1$ - and angular size greater than 11".4, to ensure all sources are resolved in FIRST. Galaxies were visually classified, and labelled with an FR class only if at least two of three human classifiers agreed on a classification.

FRICAT is comprised of a total of 219 galaxies with FRI morphology; no specific morphological labels were provided for the base catalogue, although the authors note that their inclusion criteria both permitted the presence of narrow-angle tail sources and rejected wide-angle tail sources.

FRIICAT consists of 122 galaxies with FRII morphology; again, no specific morphological labels were provided in the catalogue, but the authors state that the majority of the galaxies used were ``double" sources, i.e. ones where there is no detectable core emission, with all radio flux being in the lobes.

The criterion of requiring the agreement of multiple human classifiers for inclusion in these catalogues was commented by the authors to significantly limit the number of suitable sources within the population they examined, stating that ``more than half of the 714 radio galaxies extended more than 30\,kpc cannot be allocated to any FR class". As a result, sources from these catalogues are taken to be equivalent to MiraBest's ``confidently labelled" images, as FRICAT and FRIICAT's construction does not allow for morphological ambiguity.

A direct comparison between \cite{miraghaei2017nuclear} and these two catalogues is also made in \cite{capetti2017friicat}, as all three catalogues draw from the same parent sample of sources in \cite{best2012fundamental}. It is noted that, despite this, only around 25\% of the sources in FRICAT and FRIICAT are included in \cite{miraghaei2017nuclear}'s catalogue; this is ascribed in part to different criteria for inclusion for properties such as flux and angular size, and in part because \cite{miraghaei2017nuclear} requires that its sources be labelled as having multiple radio components, thus rejecting many FRIs without clearly separated radio emission that appear in FRICAT. On the whole, however, the authors find that, despite different selection methods, the properties derived from each set of FR galaxies are not dissimilar.

\subsection{The Combined NVSS-FIRST Galaxies sample (CoNFIG)}

CoNFIG's construction of a catalogue of Fanaroff-Riley galaxies was similarly motivated to FRICAT and FRIICAT; to provide as large as possible a dataset of labelled FR galaxies to allow study of their properties for a variety of purposes, but notably to question whether the Fanaroff-Riley dichotomy was the best method to classify these types of galaxies, given the existence of hybrid sources, which show characteristics of both \citep{gendre2008combined}.

Sources in CoNFIG, as the catalogue's name suggests, were identified amongst the population of sources visible in both NVSS and FIRST with relatively high NVSS flux density. The catalogue is comprised of four sets of combined observations (CoNFIG-1 through CoNFIG-4), and the flux density requirements differ between the different observation sets ($S_{\rm NVSS} > 1.3$, $0.8$, $0.2$, and $0.05$ Jy respectively).

In all cases, sources were morphologically classified either by identifying an established label from the 3CRR catalogue \citep{laing1983bright}, or by visual inspection of NVSS and FIRST contours. Possible labels within this dataset are FRI, FRII, ``compact" (size < 3"), and ``unclassifiable" (extended, but with morphology that cannot be assigned an FR class). It is of note that any observed hybrid sources were `` classified according to the characteristics of the most prominent jet", and thus CoNFIG contains an indeterminate quantity of hybrid sources.

The total population of CoNFIG is 71 FRIs, 466 FRIIs, 285 compact objects, and 37 unclassifiable sources. This population shows clear imbalance in favour of FRIIs, which is not unexpected; given that FRIIs have intrinsically higher luminosities than FRIs, an overrepresentation of FRIIs at high redshift may be anticipated, particularly for the CoNFIG populations with a larger minimum flux density requirement. Consequently, it is likely that a bias towards FRIIs could have been introduced. Because of this bias, CoNFIG alone is not especially well-suited for use as a FRI/FRII dataset without appreciable data augmentation being used upon its FRI sources.

\subsection{The FR-DEEP batched dataset}
\label{section: frdeep}

The FR-DEEP batched dataset \citep{tang2019transfer} is a catalogue of Fanaroff-Riley galaxies of classes FRI and FRII, drawing its sources from a combination of FRICAT and CoNFIG, and using image data from both FIRST and NVSS.

\subsubsection{Source selection}
\label{section: frdeep sources}

As the methods used to compile these two catalogues differed in their methods of classification, the CoNFIG sources were subject to additional inspection to ensure their suitability for FR-DEEP.

A spectroscopic redshift was required for inclusion in this dataset; these were not available for all CoNFIG sources, resulting in an initial reduction to 638 sources. Compact and uncertain sources were removed, as the lack of definite FR class rendered them useless for a binary FR dataset. The remaining FRI and FRII sources are labelled as either ``confirmed" or ``possible" in their classification within CoNFIG, via additional visual inspection via either the VLBA calibrator list \citep{beasley2002vlba, fomalont2003second, petrov2005third, petrov2006fourth, kovalev2007fifth} or the Pearson-Readhead survey \citep{pearson1988milliarcsecond}; only ``confirmed" sources being included in this dataset. Although the authors do not explicitly state as much, it is likely that, by rejecting sources with only ``possible" FR classification, many if not all of the hybrid sources within CoNFIG will have been removed from FR-DEEP, reducing the risk of ``confusing" morphologies being present. With this complete, a set of 50 FRIs and 390 FRIIs was deemed suitable for use.

The reduced CoNFIG sample was then combined with the 219 FRICAT FRI sources. As three of the FRIs were found in both catalogues, this resulted in a final population of 266 FRIs and 390 FRIIs. As this represents a split of approximately 40\% FRI, 60\% FRII, it is a noticeably imbalanced dataset still; \cite{tang2019transfer} acknowledge this, and suggest that FR-DEEP should be augmented so to result in a more even balance between the classes.

\subsubsection{Construction}

The data processing methods used in FR-DEEP, as briefly mentioned in Section~\ref{section: processing methods}, were used as a model for MiraBest. As such, the steps performed to prepare images for inclusion in FR-DEEP are broadly identical, and Section~\ref{section: processing methods} should be referred to for full detail on these; we will here identify principally where the methods used for FR-DEEP differ from MiraBest.

FR-DEEP was designed for transfer learning, to evaluate the performance of a classifier trained upon images from one survey when used to classify images from a different survey. Consequently, two variants were created using the same list of sources: FRDEEPF, which uses data from FIRST, and FRDEEPN, which uses data from NVSS. 

Image data for both sets of sources were initially downloaded from SkyView Virtual Observatory as $300\times300$ pixel images centred upon the FR source; this corresponds to an angular size of 540" for FIRST and 4500" (75') for NVSS. As both NVSS and FIRST images had radio background noise present, images underwent a sigma-clipping process identical to that used on MiraBest to minimise this noise, and were scaled using a likewise identical method. 

At this stage, the dataset was augmented via image rotation to create a far larger quantity of images with a more even split between classes. FRI images were rotated in 1$^\circ$ increments between 1$^\circ$ and 73$^\circ$, while FRII images were rotated between 1$^\circ$ and 50$^\circ$, resulting in approximately 20,000 images of each class. As there is no preferred orientation for FR galaxies in space,  this method of augmentation is an effective method to artificially increase the number of images that can be used in a dataset; however, it being performed during dataset construction means that any later attempts to augment the dataset via rotation have a risk of effectively cancelling this rotation out, resulting in multiple identical images being present in the doubly-augmented dataset. If any further data augmentation were deemed desirable, rotation should be avoided in favour of methods such as flipping, cropping, and scaling the sources.

Once the dataset had been augmented, all of the images were clipped down to a size of $150\times150$ pixels, again centred on the FR source. Unlike MiraBest, no circular mask was used to remove data at a radius of > 75 pixels from the image centre; this potentially allows for a slightly greater proportion of bright background sources to have been included, but as discussed in Section~\ref{section: removing background} the inclusion of such sources is not generally expected to have deleterious effects upon a dataset. The images were then converted into PNGs, and the finalised dataset was separated into a training set of 27,447 images and a test set of 11,690 images, equivalent to an approximately 70\%/30\% test-train split. As FR-DEEP was designed for binary classification, it expresses its labels as vectors: [1,0] for FRIs, and [0,1] for FRIIs. This is functionally equivalent to a MBFRFull's system of labelling FRIs as class 0 and FRIIs as class 1.

\subsubsection{Comparison to MiraBest}

In its unaugmented form, FR-DEEP consists of a total of 656 images, and hence contains around half as many sources as MiraBest, and has a noticeably more imbalanced FRI/FRII split of 40\%/60\% vs. MiraBest's 48\%/52\%. As mentioned in Section~\ref{section: frdeep sources}, augmentation was viewed by \cite{tang2019transfer} as a necessary step to ensure both a better balance between classes and to create a dataset of size comparable to other contemporary machine learning datasets. As many commonly used models are not equivariant, they will identify each rotated image as a completely unique source, and thus augmenting the dataset via rotation is not expected to introduce overfitting. However, it is nonetheless apparent that more information about the variety of morphologies in FR galaxies may be obtained from having a greater number of unique sources than by providing multiple rotated images of the same source; consequently, MiraBest may be considered to provide a broader representation of the FR population as a whole.

FR-DEEP does not account for morphological variation in its sources beyond the binary FR classes; it is unclear from FRICAT, FRIICAT and CoNFIG's source information if any sources with nonstandard morphology were included, or if all sources represent ``classic" morphology. If the former is the case, this does not necessarily make FR-DEEP less useful a dataset, as the inclusion of these sources would be expected to result in models flexible enough to be able to identify nonstandard morphology as belonging to the appropriate FR class, but it would reduce the ability to identify any characteristics that the model might associate with such subclasses without manually inspecting and relabelling all images with an appropriate subclass label. If instead the latter is the case, and all images are of ``standard" morphology, then no relabelling is necessary; however, given that all images would then be both confidently-labelled and standard-morphology, it might suffer from the effects of being overly curated - namely, resulting in models that are overconfident in their predictions on sources with morphology they have not been exposed to during training. MiraBest's explicit inclusion of uncertainly-labelled sources as well as morphological subclasses allows for direct comparison of model performance on different combinations of subclass and human labelling confidence, permitting a more in-depth study of which types of sources a classifier tends to struggle with and whether this corresponds to similar difficulty in classification by humans.

MiraBest was designed to match the construction methods of FR-DEEP closely enough that the two catalogues could potentially be merged if a yet larger dataset were desirable; besides the addition of a centred mask, MiraBest and FRDEEPF images are identically processed. Doing so, however, would require that the combined catalogue were searched for duplicate images before it were used with any machine learning model. As around 25\% of sources in \cite{miraghaei2017nuclear} were noted to also be present in FRICAT and FRIICAT (see Section~\ref{section: fricat}), so it is to be expected that multiple sources found in MiraBest may also be found in FR-DEEP. While a small quantity of duplicate images would be unlikely to significantly impact any model that trained upon this combined dataset, if as many as 25\% of FR-DEEP's images have duplicates in MiraBest then their inclusion is expected to have a negative impact upon the training process, either by providing extra sources that add no new information if both present in the training set, or by allowing for overfitting if identical images are in the training and test sets. Removing any duplicate sources, then, is considered of significant importance if combining these two datasets. On the whole, if a larger number of FR sources were deemed necessary for machine learning, one of the simplest methods would be to merge MiraBest and FR-DEEP with appropriate filtering to identify possible image duplicates, with the caveat that some effort should be made to investigate if any sources show obviously nonstandard morphology and relabel them accordingly if so.

\subsection{Aniyan \& Thorat's catalogue}

\cite{aniyan2017classifying}'s work presents a dataset of Fanaroff-Riley galaxies of classes FRI and FRII as well as radio galaxies with bent-tail morphologies, using sources from CoNFIG, FRICAT and \cite{proctor2011morphological}'s catalogue of bent radio galaxies. All image data were obtained from FIRST. The authors did not offer a name for this dataset, but for brevity it will be referred to hereon as AT17.

\subsubsection{Source selection}

AT17's methodology of initially selecting sources from CoNFIG was similar to \cite{tang2019transfer}, in that they chose to use only images which were labelled by CoNFIG as ``confident" FRI or FRII sources, rejecting all compact objects, unclassifiable sources, and sources with less certain morphological class which were labelled with a ``possible" FR class, resulting in a sample of 50 FRIs and 390 FRIIs. Likewise, to reduce the imbalance between classes, FRICAT's 219 FRI sources were added. While MiraBest treats bent-tail morphologies as simply being nonstandard morphological variants of FRI galaxies, \cite{aniyan2017classifying} chose to treat them as a third class of radio galaxy; as the other FR galaxies are presumed to show ``standard" morphology, they could be considered distinct enough to form their own population. \cite{proctor2011morphological}'s catalogue provides a number of varieties of bent-trail galaxies, but only those labelled as being both confidently classified and either wide-angle-tail or narrow-tail morphology were included in AT17, for a total of 299 bent-tail sources.

From this initial sample, sources were then inspected, and rejected for inclusion if they showed ``strong artefacts" (which are not elaborated on, but are assumed to be inherent in FIRST's data), if multiple sources were visible, or if their angular size was too great for intended image size. Once this was done, any images found to be present in both the FR sample and the bent-tail sample were removed to prevent confusion from identical sources having multiple labels, and FRICAT sources were visually inspected to remove any additional bent-tail sources. With this done, a final population of 178 FRIs, 284 FRIIs and 254 bent-tail galaxies was obtained.

\subsubsection{Construction}

The preprocessing methods used by \cite{aniyan2017classifying} were used as a basis for those of FR-DEEP; consequently, image processing follows a very similar method as has been previously discussed. Images of all sources were obtained from FIRST data at an initial size of $300\times300$ pixels, and all pixels below $3\sigma$ of the image mean were clipped. Sigma-clipping to the levels of $2\sigma$ and $5\sigma$ were also trialled, but it was found that this resulted in poor accuracy in any classifiers used; while the authors do not state as much, it is assumed that sigma-clipping to $2\sigma$ is likely to leave some radio background present, leaving machine learning models attempting to identify features in noise, while sigma-clipping to $5\sigma$ could be expected to remove portions of diffuse emission which are significant features in both FRI and bent-tail sources, destroying useful source data. No normalisation was performed; if not intending to convert an image to PNG or another similar graphics format, normalising to the colour values of said format is not necessary. However, as discussed previously, the inherent flux density properties of FR galaxies are such that foregoing normalisation might lead to a machine learning model largely ignoring morphological features to classify based on maximum flux density; this possibility is not discussed by the authors, so this effect may not be clearly apparent in their work.

In the sample used to create this dataset, FRIs are noticeably underrepresented compared to the other two classes; as with FR-DEEP, the dataset was augmented via rotation, with each image rotated multiple times to produce an approximately even balance of classes, but these augmentations were only to be applied to the training set, with the validation set left unaltered. Consequently, the train-validation split was applied at this stage; 53 FRIs, 57 FRIIs and 77 bent-tail sources (corresponding to around 30\% of the total dataset population) were selected for the validation set and excluded from the augmentation process.

Unlike FR-DEEP, the increments at which each training set image was rotated were varied based on class rather than setting differing maximum angles of rotation; FRIs were rotated at increments of 1$^\circ$, FRIIs at 2$^\circ$, and bent-tail sources at 3$^\circ$. The authors also state that versions of the images that had been both flipped and rotated were produced, but do not specify whether these were used in the dataset. Once the training set had been supplemented, all images were again cropped to a final size of $150\times150$ pixels.

At this stage, the training set was separated into two portions - a training set and a test set - at proportions of 80\%/20\%. By using this test set during training, the validation set separated earlier could serve as examples of truly unseen data, acting as a second check against overfitting if necessary. Following this, the final dataset population was approximately 94000 sources in the training set ($\sim$36000 FRI, $\sim$33000 FRII, $\sim$25000 bent-tail), approximately 23000 sources in the test set ($\sim$9000 FRI, $\sim$7900 FRII, $\sim$6400 bent-tail), and 187 sources in the validation set (53 FRIs, 57 FRIIs, 77 bent-tail).

\subsubsection{Comparison to MiraBest}

In its unaugmented form, AT17 contains a smaller quantity of labelled FRI and FRII sources than both FR-DEEP and MiraBest; its total of 462 FR sources makes it approximately one third the size of MiraBest, and it has a similar class balance to FR-DEEP at 39\% FRI/61\% FRII. While on the whole its image selection and processing methods are similar to the other two datasets (besides its lack of normalisation), one aspect is significantly different: the treatment of bent-tail sources as a third class.

At present, the exact nature of bent-tail sources is still not completely certain. Some view them as FRI galaxies with unusual morphologies; \cite{miraghaei2017nuclear} are among this group, including wide-angle-tail sources in their catalogue with a FRI label, as do \cite{gregory2017narrow}. Others view them as potentially being a completely separate population, given that their bent morphology often seems more alike FRIIs than FRIs; while constructing WATCAT (a wide-angle-tail galaxy catalogue built to complement FRICAT and FRIICAT), \cite{missaglia2019watcat} found that at 1.4\,GHz wide-angle-tail sources have flux density more comparable to FRIIs, above the classic flux cut-off, but when their radio power is plotted against optical magnitude they instead fall within the region populated by FRIs. \cite{aniyan2017classifying} find that their three-class classifier is much more likely to mislabel bent-tail sources as FRIIs than FRIs, although as discussed previously this may have been affected by the lack of image normalisation. 

While there does not yet appear to be a full consensus as to where bent-tail galaxies fit in to the FR dichotomy, it is generally agreed that if not an entirely separate class, bent-tail sources resemble FRIs more closely than either FRII or hybrid sources. Consequently, if we are to treat bent-tail sources as a subset of FRI galaxies (as done within MiraBest), AT17 becomes imbalanced in the opposite direction: over 60\% of the unaugmented images are of FRIs, with ~60\% of that group being bent-tail sources, which does not reflect the true prevalence of this morphology. Because of this imbalance and focus upon only standard FRI, standard FRII, and bent-tail morphologies, AT17 may be considered to be a less diverse catalogue in terms of overall FR morphology than MiraBest, in addition to containing a noticeably smaller population of sources.

Despite sharing broadly the same image processing methods, MiraBest and AT17 are not immediately compatible to be merged into a single dataset because of AT17's lack of scaling and the previously discussed mismatch in their class structure. The former is more readily overcome, as the application of the same scaling method used in both MiraBest and FR-DEEP is still possible to perform on the processed AT17 images, while the latter requires consideration as to the most recent consensus on the properties of bent-tail galaxies. Additionally, as with FR-DEEP, the presence of duplicate sources is a concern; as previously discussed, \cite{miraghaei2017nuclear}'s catalogue shares a noticeable population of sources with FRICAT and FRIICAT, and while it is unclear whether any of the bent-tail sources of \cite{proctor2011morphological}'s are likewise present, both catalogues drawing from data from the FIRST survey means it is entirely possible. Combining these catalogues, then, is not trivial, and this difficulty makes AT17 a less appealing immediate candidate than FR-DEEP for expanding upon the overall source count; however, if a larger population of bent-tail sources is desired for study, AT17 is the most straightforward catalogue to adapt for use alongside MiraBest.

\section{The MiraBest Dataset Python Class}
\label{sec:usage}

The structure of the MiraBest dataset mimicks that of the widely used MNIST \citep{lecunmnist} and (e.g.) CIFAR \citep{cifar10} datasets that are used with popular deep learning software packages such as Pytorch \citep{pytorch} and Keras \citep{chollet2015keras}. 

A Python class is provided with the dataset itself in order to facilitate its use with these packages with a structure inherited from the Pytorch {\tt data.Dataset} class. When using this class there is no need to download the MiraBest dataset independently as the data loader will pull a remote copy automatically if no local instance is found. As for the Pytorch MNIST and CIFAR data loaders a checksum is instituted in order to avoid corrupted versions of the dataset being created. 

The metadata for each sample includes both a ``label" and a ``fine label", where the label indicates the binary FRI/FRII classification, see Table~\ref{tab: mirabest derivatives}, and the fine label indicates the morphological sub-classification, see Table~\ref{tab: mirabest structure}. Child classes for the sub-samples MiraBest Confident, MiraBest Uncertain and MiraBest Hybrid, see Table~\ref{tab: mirabest derivatives}, are also included in the Dataset class. 

\subsection{Dataset normalisation}
\label{sec:normalise}

For deep learning applications it is standard practice to normalise individual data samples from the data set by shifting and scaling as a function of the mean and variance calculated from the full training set \citep{tricks}. The normalisation parameters for the MiraBest training dataset and its derivative training datasets are listed in Table~\ref{tab:norms}.

\begin{table}
    \centering
    \caption{Mean and standard deviation of the MiraBest training dataset and its derivatives.}
    \label{tab:norms}
    \begin{tabular}{lcc}
        \hline
        Dataset & $\mu$ & $\sigma$ \\\hline
        MiraBest (full) & 0.0031 & 0.0352\\
        MiraBest Confident & 0.0031 & 0.0350\\
        MiraBest Uncertain & 0.0031 & 0.0356\\
        MiraBest Hybrid & 0.0036 & 0.0375\\
        CRUMB & 0.0029 & 0.0341\\\hline
    \end{tabular}
\end{table}

\section{CRUMB: Collected Radiogalaxies Using MiraBest}

All of the FR galaxy datasets discussed in this work use image data from the VLA FIRST survey, and as such it is possible to combine them into a single dataset without encountering problems as a result of differing survey properties, such as (e.g.) angular resolution. Doing this would allow for a larger number of sources to be used for training and, if the parent dataset of each source were to be labelled, allow for direct comparison of performance on each of these datasets. However, this cannot be done by simply merging the datasets together; both FR-DEEP and AT17 draw from the same catalogues for their sources, meaning that a simple merge would not only result in duplications of sources but potentially those duplications having multiple different labels, which would effectively result in label noise. 
This motivated the creation of the Collected Radiogalaxies Using MiraBest (CRUMB) dataset, which is a cross-matched combination of MiraBest, FR-DEEP, AT17 and MiraBest Hybrid. This dataset retains not only a record of which parent datasets each source can be found in but their label in each of these datasets, and hence offers the ability to select labels from the user's catalogue of choice.

\subsection{Constructing CRUMB}

To construct CRUMB, the full lists of the sources used in FR-DEEP \citep{tang2019frdeep} and AT17 \citep{aniyan2017classifying} were cross-matched with the sources in MiraBest and MiraBest Hybrid to identify duplicates. Since the location of the source centre was not expected to be exactly consistent between different catalogues, duplicates were found by searching for sources which had coordinates within 270 arcseconds of ones another, i.e. within the same image using MiraBest's image size, and checking whether the coordinates aligned with the same source using visual inspection. Using this method, a total of 2100 unique sources were found to exist when combining these four datasets, with 541 being present in more than one dataset.

For this combined sourcelist, the class labels of sources which appear in more than one dataset were examined to check for disagreements in FR class between different datasets. The vast majority of the duplicate sources (518 of 541) were found to have been classified with the same overall FR class in all datasets, with 470 also agreeing on morphological subclass. A small number of sources (15) showed clearly contradictory labels, with sources labelled as FRI in one dataset being labelled as FRII in another. This demonstrates the label noise issue that can arise if merging machine learning catalogues without checking for duplicate sources, which if not addressed would result in models learning multiple labels for the same source.

To allow for this ambiguity in class to be retained, CRUMB uses a labelling system which provides both a ``basic" and a ``complete" label. The ``complete" label is represented by vector with four entries, each of which represents a source's class label in each of the four parent datasets as shown in Table~\ref{tab: crumb structure}. If a source is not present in a given dataset, it is denoted with ``-1" in the relevant entry. This allows for multiple  class labels to be registered; for example, a vector of [0, -1, 2, -1] would correspond to a source which is labelled as a confident standard-morphology FRI in MiraBest, a bent source in AT17, and is not present in FR-DEEP or MiraBest Hybrid.

\begin{table}
    \centering
    \caption{The ``complete" class labels used within the CRUMB dataset. If a source is not present in a dataset, it is labelled as ``-1".}
    \label{tab: crumb structure}
    \begin{tabular}{llll}
        \hline
        Entry 0 & Entry 1 & Entry 2 & Entry 3\\
        MiraBest & FR-DEEP & AT17 & MB Hybrid \\\hline
        0 - 9 & 0 - FRI & 0 - FRI & 0 - Conf. hybrid\\
        (see Table~\ref{tab: mirabest structure}) & 1 - FRII & 1 - FRII & 1 - Unc. hybrid\\
        & & 2 - Bent &\\
    \end{tabular}
\end{table}

Additionally, each source is labelled with one of three ``basic" labels: FRI (0), FRII (1) and hybrid (3). These labels are assigned by the majority label in all the datasets a source appears in; in the case of two contradictory labels, we favour the label provided by MiraBest. Using this method, all sources labelled by AT17 as ``bent" are folded into the FRI class, and a total of 1009 FRIs, 998 FRIIs and 97 hybrid sources are included in CRUMB.

Images in the CRUMB dataset were processed in the same manner as for MiraBest. Because of the ambiguity in ``true" label and lack of information on redshift and angular size for many sources, image file names are formatted as ``[source right ascension]\_[source declination]" for consistency across this combined dataset. Both the file name and complete label may be retrieved for any source using the built-in ``filenames" and ``complete\_labels" methods.

\section{Use of MiraBest in the literature}
\label{sec:literature}

The first use of MiraBest was made by \cite{bowles2021attention} who demonstrated that an attention-gated CNN model recovered an $92\%$ accuracy on the MiraBest Confident test set ($84\%$ accuracy on the full MiraBest test set including uncertain samples), exceeding the $88\%$ classification accuracy attained using the FRDEEP-F dataset by \cite{tang2019transfer}. \cite{e2cnn} used the MiraBest dataset to demonstrate that classification performance is modestly improved by enforcing both cyclic and dihedral equivariance in the convolution kernels of a CNN for FR classification and that E(2)-equivariant models were able to reduce variations in model confidence as a function of galaxy orientation. \cite{inigo2022} explored the effect of dataset shift in semi-supervised learning (SSL) by combining labelled data from MiraBest with a larger unlabelled data pool from the Radio Galaxy Zoo catalogue (Wong et al. in prep), demonstrating that when different underlying catalogues drawn from the same radio survey are used to provide the labelled and unlabelled data sets required for SSL, a significant drop in classification performance is observed. In \cite{mohan2022} the uncertainty associated with classification of individual data samples within the MiraBest dataset was explored using Bayesian deep learning, confirming that the machine learning model was less confident about the samples qualified as Uncertain by the MiraBest labelling scheme than those labelled as Confident, and that this was amplified for samples labelled as Hybrid.

\section{Conclusions}
\label{sec:conclusions}

Machine learning datasets of astronomical data often have different requirements than astronomical datasets used for other purposes. For image classification, principle amongst these requirements is having reliably-labelled data that either exists in large enough quantities to not necessitate augmentation, or exists in smaller quantities that may be augmented in such a way that the dataset size can be artificially increased without resulting in overfitting. These requirements often result in astronomical ML datasets being created for the specific research needs of a small number of individuals and not being made readily available for broader use, as it is assumed that others wishing to construct a similar dataset will likewise independently seek out suitable sources which meet their needs.

Fanaroff-Riley galaxy classification has previously been performed using machine learning, but the majority of existing catalogues of FR galaxies have not been used to produce publicly accessible image datasets. Datasets which have been made accessible, such as FR-DEEP, were found to be limited to only binary FR sources and to contain fewer images - $\mathcal{O}(10^2)$ - than the largest current catalogues of FR galaxies, which consist of $\mathcal{O}(10^3)$ examples. Because of this, we felt the need to produce a new publicly accessible machine learning dataset for Fanaroff-Riley galaxies, and created MiraBest for this purpose. 

At time of writing, MiraBest is believed to be the largest publicly available image dataset labelled according to the FR classification and most diverse in terms of inclusion of rarer morphologies. Additionally, the option of including the more morphologically ambiguous data represented by the "uncertainly-labelled" images means that MiraBest may be considered a less curated dataset than many other image classification datasets, which largely present only clear and unambiguous images of the target classes. Because of this, MiraBest is suitable for examining the ability of classifiers to identify unusual and ambiguous sources, and whether the inclusion of these sources in a model's training data helps or hinders performance both on the whole and in ability to recognise these unusual sources in particular. 

\section*{Acknowledgements}

FP gratefully acknowledges support from STFC and IBM through the iCASE studentship ST/P006795/1. AMS gratefully acknowledges support from an Alan Turing Institute AI Fellowship EP/V030302/1. The authors would also like to thank Hongming Tang and Kshitij Thorat for their assistance in providing the source lists used in their datasets.

\section*{Data Availability}

MiraBest has been made accessible for public download via the Zenodo website (\url{https://doi.org/10.5281/zenodo.4288837}; DOI: 10.5281/zenodo.4288837), allowing it to be used for any research applications using FR galaxies. Information about its construction has also been provided, permitting its integration with other datasets if desired and making it possible to supplement it with additional FR galaxies processed in an identical format if other large catalogues of these galaxies should become available in the future. CRUMB has likewise been made accessible for public download via Zenodo (\url{https://doi.org/10.5281/zenodo.7948346}, DOI: 10.5281/zenodo.7948346).



\bibliographystyle{mnras}
\bibliography{mirabest_paper} 








\bsp	
\label{lastpage}
\end{document}